# A Dual Approach for Solving Nonlinear Infinite-Norm Minimization Problems with Applications in Separable Cases

**Wajeb GHARIBI**[1]     **Yong XIA**[2,3]

[1] College of Computer Science, King Khalid University, Abha, Kingdom of Saudi Arabia.

[2] State Key Laboratory of Scientific/Engineering Computing, Institute of Computational Mathematics and Scientific/Engineering Computing, The Academy of Mathematics and Systems Sciences, Chinese Academy of Sciences, P. O. Box 2719, Beijing, 100080, P. R. China,

[3] Graduate School of the Chinese Academy of Sciences, Beijing, 100080, P. R. China

## ABSTRACT

In this paper, we focus on nonlinear infinite-norm minimization problems that have many applications, especially in computer science and operations research. We set a reliable Lagrangian dual aproach for solving this kind of problems in general, and based on this method, we propose an algorithm for the mixed linear and nonlinear infinite-norm minimization cases with numerical results.

# 1. Introduction

In this paper, we focus on solving nonlinear infinite-norm minimization problems with applications in separable cases.

Consider the following problem; model of nonlinear functions that can depend on multiple parameters:

(1.1)
$$\min_{y \in R^n} \|F(y)\|_\infty$$

where $F: R^n \to R^m$ (m>n) and $y \in R^n$ is a vector of variables ($\|x\|_\infty = \max_{1 \leq i \leq n} |x_i|$).

Generally, the problem (1.1) is difficult to be solved because of both the nonlinearity of F and the nondifferentiality of the infinity-norm.

The usual approach [1] to deal with (1.1) is using p-norm to approximate infinite norm due to the equality $\|F\|_\infty = \lim_{p \to \infty} \|F\|_p$, where $\|x\|_p = \left( \sum_{i=1}^n |x_i|^p \right)^{\frac{1}{p}}$.

In this article, we will give a dual approach to handle this problem

As a sepcial case of (1.1), a model of a linear combination of nonlinear functions that can depend on multiple parameters:

(1.2)
$$\min_{\substack{x \in R^{n_1} \\ y \in R^{n_2}}} \|A(y)x - b(y)\|_\infty,$$

where, $A(y) \in R^{m \times n_1}$, $b(y) \in R^m$, ( generally $m > n_1 + n_2$ ), are nonlinear.

This type of problems is very common and has a wide range of applications in different areas, such as inverse problems, signal analysis, mechanical systems, neural networks, communications, robotics and vision, electrical engineering, environmental sciences, to name just a few [2].

Recently, Golub and Pereyra [2] have proposed an algorithm for solving the least squares problem:

(1.3) $$\min_{\substack{y \in R^p \\ z \in R^q}} \|F(z)y - g(z)\|_2$$

where, $F \in R^{m \times p}$ is a variable matrix and $g \in R^m$ is a variable vector.

Obviously, the problem (1.1) is more difficult than the Problem (1.3) and it has some interesting specifications.

In this paper, we study the problem (1.1) and set an algorithm for finding an optimal solution provided with computational results.

In general, our problem derived from the special nonlinear data fitting problem:

(1.4) $$\min_{\substack{a \in R^p \\ b \in R^q}} \frac{1}{2} \sum_{i=1}^{m} [y_i - \sum_{j=1}^{p} a_j \phi_j(b, t_i)]^2$$

(where $\phi_j(b, t_i): R^p \to R;\ p + q = n$), which has the following standard form:

(1.5) $$\min_{\substack{a \in R^p \\ b \in R^q}} \psi(a, b) = \frac{1}{2} \|y(b) - \phi(b)a\|_2^2$$

where $y(b) \in R^m;\ \phi(b) \in R^{m \times p}$, with $(\phi(b))_{ij} = \phi_j(b, t_i);\ i = 1, 2, ..., m,\ j = 1, 2, ..., p.$

Golub and Pereyra [3] proposed a variable projected method for solving this type of problems with the 2-norm.

Our paper is organized as follows. In the next section, we present dual approach for solving the general case of our problem. In section 3, an algorithm for finding solution of a special case of problem (1.1) is presented. Numerical results and discussion are reported in section 4. Conclusion is made in the last section.

## 2. A Dual Approach

Consider the problem (1.1)

$$\min_{y \in R^n} \|F(y)\|_\infty$$

Which can be written as:

$$\min z$$
$$s.t. \quad z \geq |F_i(y)|, \quad i = 1, 2, ..., m$$

and this is equivalent to the problem:

$$\min z$$
$$s.t. \quad z^2 \geq (F_i(y))^2, \quad i = 1, 2, ..., m$$

It is also equivalent to

(2.1)
$$\min z^2$$
$$s.t. \quad z^2 \geq (F_i(y))^2, \quad i = 1, 2, ..., m$$

because of the following proposition.

**Proposition 2.1.** The problem $\min_{x \in C} f(x)$ is equivalent to $\min_{x \in C} g(f(x))$ for any monotone function $g : S \to R$, where $S := \{f(x) : x \in C\}$.

Moreover, (2.1) is equivalent to

(2.2)
$$\min t$$
$$s.t. \quad t \geq (F_i(y))^2, \quad i = 1, 2, ..., m$$

Now, consider the Lagrangian dual problem of (2.2) which has the form

(2.3)
$$\max_{\substack{\lambda \geq 0 \\ \sum_i \lambda_i = 1}} \min_y \sum_{i=1}^m \lambda_i F_i(y)^2$$

Note that, for any fixed $\lambda$ the inner subproblem of (2.3) becomes nonlinear least squares problem which has been well studied [3-6].

Due to the weak duality theory, we have

**Proposition 2.2.** The optimal objective function value of Problem (2.2) is greater than or equal to that of Problem (2.3).

**Proposition 2.3.** The optimal objective function values of problems (2.2) and (2.3) are equal when one of the following conditions is satified

(a) $F$ is a linear function for $i = 1, 2, ..., m$,

(b) $F_i$ is a convex function and nonnegative for $i = 1, 2, ..., m$,

(c) $F_i$ is a concave function and nonpositive for $i = 1, 2, ..., m$.

There are some common approaches for solving the problem (2.3), such as the subgradient method. But it is difficult to solve its inner minimal problem to optimality in case when $F_i$ does not satisfy any of the previous conditions shown in Proposition 2.2. Usually, only local solutions can be obtained. Here, keep in mind that we really want to solve the problem (2.2), and due to Proposition 2.1, we do not need to exactly solve the problem (2.3); local solutions seem to be better ones.

## 3. An Algorithm For Special Case

In this section, we set an algorithm for solving the problem (1.2)

$$\min_{\substack{x \in R^{n_1} \\ y \in R^{n_2}}} \|A(y)x - b(y)\|_\infty$$

where, $A(y) \in R^{m \times n_1}$, $b(y) \in R^m$, ( generally $m > n_1 + n_2$ ), are nonlinear.

**Theorem 3.1.** If $(x^*, y^*)$ is the optimal solution of the problem (1.2) then

$$\min_{x \in R^{m \times n_1}} \|A(y^*)x - b(y*)\|_\infty$$

is equivalent to the linear program:

(3.1)
$$\begin{aligned}
&\min z \\
&s.t. \ z \geq (A(y^*)x - b(y^*))_i, \quad i = 1, 2, ..., m \\
&\quad\ \ z \geq -(A(y^*)x - b(y^*))_i, \quad i = 1, 2, ..., m
\end{aligned}$$

Now, denote the optimal solution of (3.1) by $x^{**}$, then $(x^{**}, y^*)$ is also the optimal solution of the problem (1.2).

Following Theorem 3.1, we can set the coming alternate algorithm.

**Algorithm 3.1.**

**Step 1** Set initial value $(x_0, y_0)$.

**Step 2** Solve the following linear problem for fixed $y_0$

$$\min_{x \in R^{n_1}} \|A(y_0)x - b(y_0)\|_\infty$$

and obtain the optimal solution $x_1$

**Step 3** Solve the following problem for fixed $x_1$

(3.2)
$$\min_{y \in R^m} \|A(y)x_1 - b(y)\|_\infty$$

and obtain the optimal solution $y_1$.

**Step 4** Do the line search

(3.3)
$$\min_{\beta \in R^m} \|A(y_0 + \beta(y_1 - y_0))(x_0 + \beta(x_1 - x_0)) - b((y_0 + \beta(y_1 - y_0)))\|_\infty$$

and obtain the optimal solution $\beta^*$.

**Step 5** If the stop conditions satisfied then stop. Otherwise, update $x_0 := x_0 + \beta^*(x_1 - x_0)$, $y_0 := y_0 + \beta^*(y_1 - y_0)$ and go to Step 2

Note that (3.2) becomes a general infinite-norm minimization problem with smaller dimension than that of (1.2).

## 4. Numerical Results

In this section, we implement the Algorithm 3.1 by MATLAB 6.5 using a CPU Pentium IV with 2.4 GHz. We compare two methods for solving the subproblem (3.2), iterative 2p-norm (p=1,2,...) approximation [1] and also the dual approach. Here, the dual problem is solved by the subgradient method. We call the MATLAB function FMINUNC to solve 2p-norm subproblems and LSQNONLIN to solve least squares subproblems in the dual approach. The algorithms stop when the variance between the current and the next optimal objective function is less than 1e-4.

The data of the examples are produced at random with zero optimal objective values. The dimension is set m=100. We run each algorithm 10 times independly and list the average objective value found and the average CPU time in seconds.

**Example 4. 1.** In this eample, we give fiting data for the model [3, 4]:

$$a_1 + a_2 e^{-\alpha_1 t} + a_3 e^{-\alpha_2 t}$$

The results for this problem are given in Table 1.

|  | 2p-norm | Dual approach |
|---|---|---|
| Average Optimal Objective found | 0.0027 | 0.0003 |
| Average Time in Seconds | 9.4 | 1.8 |

**Table 1**

**Example 4. 2.** The second example is given for the model [3, 4]:

$$a_1 e^{-\alpha_1 t} + a_2 e^{-\alpha_2 (t-\alpha_3)^2} + a_3 e^{-\alpha_4 (t-\alpha_5)^2} + a_4 e^{-\alpha_6 (t-\alpha_7)^2}$$

The results for this problem are given in Table 2.

|  | 2p-norm | Dual approach |
|---|---|---|
| Average Optimal Objective found | 0.0176 | 0.0024 |
| Average Time in Seconds | 11.5 | 5.2 |

**Table 2.**

**Example 4. 3.** The second example has the model [3, 4]:

$$a_1 + a_2 t + a_3 t^2 - a_4 [\frac{1}{1+((\alpha_1 + 0.5\alpha_2 - t)/\alpha_3)^2} + \frac{1}{1+((\alpha_1 - 0.5\alpha_2 - t)/\alpha_3)^2}]$$
$$-a_5 [\frac{1}{1+((\alpha_4 + 0.5\alpha_5 - t)/\alpha_6)^2} + \frac{1}{1+((\alpha_4 - 0.5\alpha_5 - t)/\alpha_6)^2}]$$
$$-a_6 [\frac{1}{1+((\alpha_7 - t)/\alpha_8)^2}]$$

The results for this problem are given in Table 3.

|  | **2p-norm** | **Dual approach** |
|---|---|---|
| Average Optimal Objective found | 0.0632 | 0.0003 |
| Average Time in Seconds | 7.2 | 0.1 |

**Table 3**

In our numerical experiments, the MATLAB function FMINUNC usually terminated until the Maximum number of function evaluations exceeded. This is partly because the 2p-norm minimization problems are difficult to be solved for large p. While our dual approach always works successfully.

## 5. Conclusion

In this paper, we proposed a dual aproach with algorithm for solving nonlinear infinite-norm minimization problems with applications in separable cases. Comparing to the 2p-norm approximation method, our dual approach works well and often gives better solutions in less time.

**Acknowledgements**

The authors would like to thank professor Ya-xiang Yuan for his generous hospitality and supervision. The first author gratefully thanks the CAS-TWAS for their kind help, support and cooperation and also wants to mention that this work done in the Institute of Computational Mathematics, Chinese Academy of

Sciences by the grant from the Award of 2005 CAS-TWAS Visting Scalar Felloship.## References

1. E. W. Cheney, Introduction to sppproximation theory. McGraw-Hill Book. Co., 1966.

2. G. Golub and V. Pereyra, Separable nonlinear least squares: The variable projection method and its applications. Inverse Problems, 19, 1-26, 2002.

3. G. H. Golub and V. Pereyra, The differentiation of pseudo-inverses and nonlinear least. squares problems whose variables separate. SIAM Journal on Numerical Analysis, Vol. 10, No. 2. (Apr., 1973), pp. 413-432.

4. L. Kaufman, A variable projected method for solving separable nonlinear least squares problems. BIT 15 (1975), 49-57.

5. L. Kaufman and V. Pereyra, A method for nonlinear least squares problems with separable nonlinear equality constraints. SIAM Journal on Numerical Analysis, Vol. 15, No. 1. (Feb., 1979), pp.12-20.

6. Axcel Ruhe and Per Ake Wedin. Algorithms for nonlinear least squares problems. SIAM Review, Vol. 22, No.3 (Jul., 1980), pp. 318-337.
10